\documentclass[fleqn,usenatbib]{mnras}

\pdfoutput=1
\usepackage{graphicx,subfigure,bm,color,psfrag,hyperref}
\usepackage{amsfonts}
\usepackage{lipsum}
\usepackage{mathtools}
\usepackage{verbatim}
\usepackage{color}
\usepackage{newtxtext,newtxmath}
\usepackage[T1]{fontenc}
\usepackage{amsmath} 

\DeclareRobustCommand{\VAN}[3]{#2}
\let\VANthebibliography\thebibliography
\def\thebibliography{\DeclareRobustCommand{\VAN}[3]{##3}\VANthebibliography}

\begin{document}

\title[A combined analysis of the $H_0$ late time direct measurements]{A combined analysis of the $H_0$ late time direct measurements and the impact on the Dark Energy sector}


\author[Eleonora Di Valentino]{Eleonora Di Valentino$^{1}$\thanks{E-mail: eleonora.di-valentino@durham.ac.uk}\\
$^{1}$Institute for Particle Physics Phenomenology, Department of Physics, Durham University, Durham, DH1 3LE, UK.}

\date{Accepted XXX. Received YYY; in original form ZZZ}

\pubyear{2021}

\label{firstpage}
\pagerange{\pageref{firstpage}--\pageref{lastpage}}
\maketitle
\begin{abstract}
We combine 23 Hubble constant measurements based on Cepheids-SN Ia, TRGB-SN Ia, Miras-SN Ia, Masers, Tully Fisher, Surface Brightness Fluctuations, SN II, Time-delay Lensing, Standard Sirens and $\gamma$-ray Attenuation, obtaining our best {\it optimistic} $H_0$ estimate, that is $H_0=72.94\pm0.75$ km/s/Mpc at 68\% CL. This is in $5.9\sigma$ tension with the $\Lambda$CDM model, therefore we evaluate its impact on the extended Dark Energy cosmological models that can alleviate the tension. We find more than $4.9\sigma$ evidence for a phantom Dark Energy equation of state in the $w$CDM scenario, the cosmological constant ruled out at more than $3\sigma$ in a $w_0w_a$CDM model and more than $5.7\sigma$ evidence for a coupling between Dark Matter and Dark Energy in the IDE scenario. Finally, we check the robustness of our results, and we quote two additional combinations of the Hubble constant. The {\it ultra-conservative} estimate, $H_0=72.7\pm 1.1$ km/s/Mpc at 68\% CL, is obtained removing the Cepheids-SN Ia and the Time-Delay Lensing based measurements, and confirms the evidence for new physics.
\end{abstract}


\section{Introduction}

The $\Lambda$CDM model provides a wonderful explanation for most of the current cosmological probes. However, its validity is questioned by the robust tensions emerged between the early and the late time Universe measurements (see~\cite{DiValentino:2020zio,DiValentino:2020vvd} for a recent overview).
In particular, statistically significant is the long standing Hubble constant tension at more than $4.4\sigma$ between the $H_0$ value estimated by Planck in~\cite{Aghanim:2018eyx}, and that measured by the SH0ES collaboration in~\cite{Riess:2019cxk} (R19). 
The tension is made even more intriguing by the several early and late time cosmological probes (see~\cite{Verde:2019ivm,Riess:2020sih}) in agreement, respectively, with Planck or R19, that make the systematic errors explanation more difficult, because biased always in the same direction.\footnote{See also~\cite{Dhawan:2017ywl,Dhawan:2020xmp} for a discussion about the effect of possible systematic errors coming from dark energy model assumptions and the wavelength region of the observations.}

A gigantic effort is put into resolving the Hubble constant tension, and many cosmological scenarios, in alternative or extending the $\Lambda$CDM model, have been considered.
We have, for example, early modifications of the expansion history, promising because, in principle, they could solve at the same time the $H_0$ tension and give a lower sound horizon $r_{\mathrm{drag}}$ at the drag epoch (\cite{Knox:2019rjx,Evslin:2017qdn}) as preferred by the Baryon Acoustic Oscillations (BAO) data. The Early Dark Energy (\cite{Pettorino:2013ia,Poulin:2018cxd,Karwal:2016vyq,Sakstein:2019fmf,Niedermann:2019olb,Akarsu:2019hmw,Ye:2020btb,Agrawal:2019lmo,Lin:2019qug,Berghaus:2019cls,Smith:2019ihp,Lucca:2020fgp,Braglia:2020bym}), and extra relativistic species at recombination (\cite{Anchordoqui:2011nh,Jacques:2013xr,Weinberg:2013kea,Anchordoqui:2012qu,Carneiro:2018xwq,Paul:2018njm,DiValentino:2015sam,Green:2019glg,Ferreira:2018vjj,Gelmini:2019deq,DiValentino:2015wba,Poulin:2018dzj,Baumann:2016wac,Barenboim:2016lxv,Zeng:2018pcv,Allahverdi:2014ppa,Braglia:2020iik}) are the most famous models, but they can not increase the Hubble constant enough to solve the tension with R19 below $3\sigma$ (\cite{Arendse:2019hev}). The late time modifications of the expansion history, instead, have the capability of completely solving the Hubble tension with R19 within $1\sigma$, but leave the sound horizon unaltered, introducing a tension with the BAO data. In this category we find the phantom Dark Energy (\cite{Aghanim:2018eyx,Yang:2018qmz,Yang:2018prh,DiValentino:2019dzu,Vagnozzi:2019ezj,DiValentino:2020naf,Keeley:2019esp,Joudaki:2016kym,Alestas:2020mvb}) and the Phenomenologically Emergent Dark Energy (\cite{Li:2019yem,Pan:2019hac,Rezaei:2020mrj,Liu:2020vgn,Li:2020ybr,Yang:2020tax}). Another promising possibility is an interaction between the Dark Matter and the Dark Energy (IDE) models (\cite{Pettorino:2013oxa,Wang:2016lxa,Kumar:2016zpg,DiValentino:2017iww,Kumar:2017dnp,vandeBruck:2017idm,Yang:2018euj,Yang:2018uae,Yang:2019uzo,Martinelli:2019dau, DiValentino:2019ffd,DiValentino:2019jae,Benevento:2020fev,Gomez-Valent:2020mqn,Lucca:2020zjb,Yang:2020uga,Yang:2019uog,Yang:2018ubt,Agrawal:2019dlm,Anchordoqui:2019amx,Anchordoqui:2020sqo}), that can solve completely the Hubble constant tension. In these models there is a flux of energy between the dark matter and the dark energy, therefore, lowering the matter density, we can have a larger $H_0$ value for the geometrical degeneracy present in the CMB data (\cite{DiValentino:2020leo}).

In this paper we combine, in an optimistic way, most of the late time measurements of the Hubble constant together, and we use this our best $H_0$ estimate to constrain some of the DE models that better solve the $H_0$ tension, namely $w$CDM, $w_0w_a$CDM and IDE. Moreover, we test the robustness of our results using two additional Hubble constant estimates, that we will call {\it conservative} and {\it ultra-conservative}. 

We introduce in Section~\ref{Observations} the data used in this paper and we explain the way we combine the different late time Hubble constant measurements, we present in Section~\ref{models} the Dark Energy models we consider in this work, we describe in Section~\ref{Method} the method used to analyse the cosmological parameters, we discuss in Section~\ref{results} the results we obtain, and we derive in Section~\ref{conclu} our conclusions.

\begin{figure*}
\includegraphics[width=\textwidth]{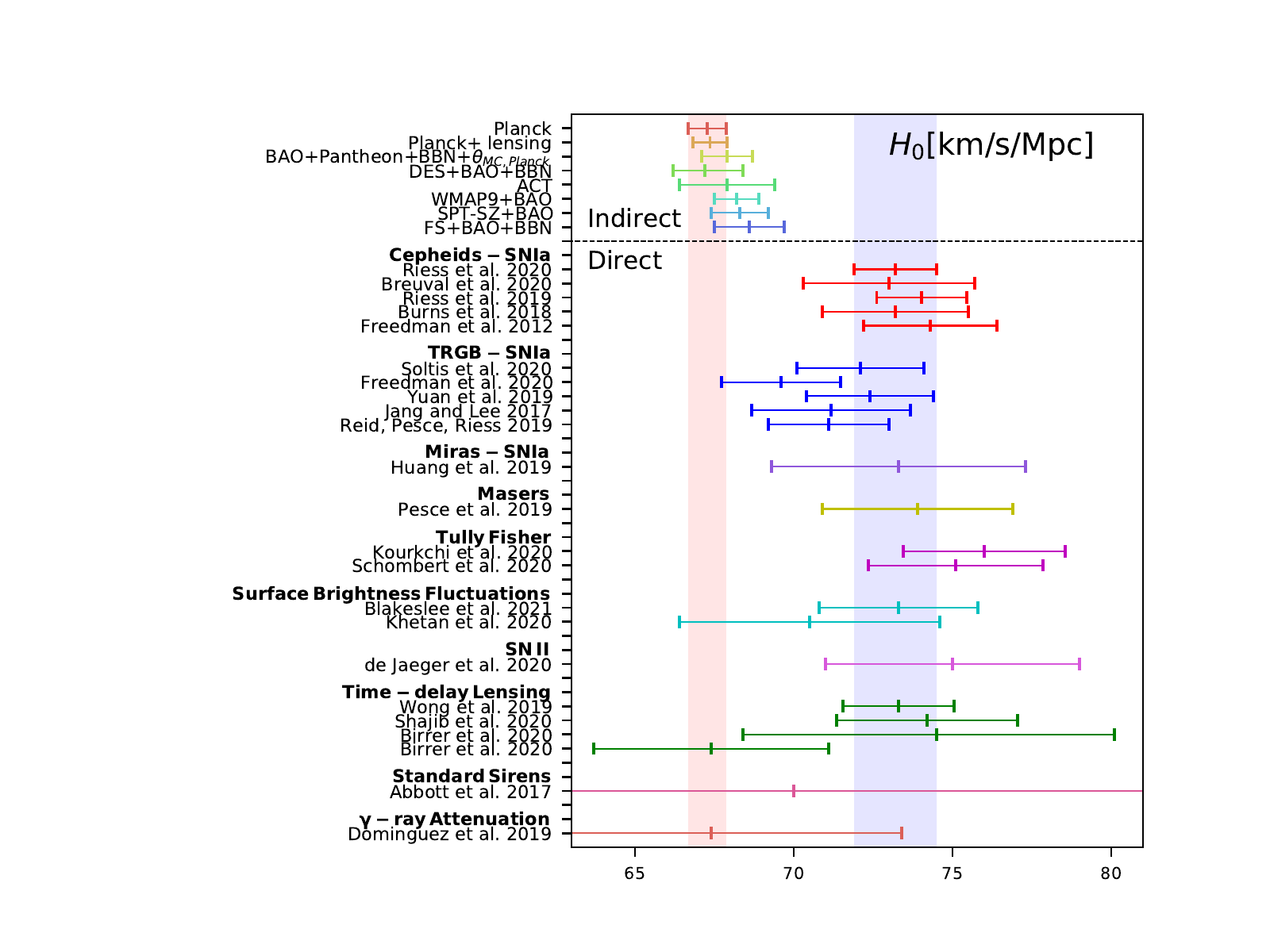}
\caption{Some of the Hubble constant measurements present in the literature: see Table~\ref{listH0} for the values. Those averaged to obtain our $H_0$ estimates are instead listed in Section~\ref{Method}.}
\label{fig-H0}
\end{figure*}

\begin{table}
\begin{center}
\begin{tabular}{|c|c|c|}
\hline
Dataset                    & $H_0$ [km/s/Mpc] \\
\hline
Planck \cite{Aghanim:2018eyx} & $67.27 \pm 0.60$ \\
Planck+lensing  \cite{Aghanim:2018eyx}  & $67.36 \pm 0.54$ \\
BAO+Pantheon+BBN+$\theta_{\rm MC, Planck}$ \cite{Aghanim:2018eyx}   & $67.9 \pm 0.8$ \\
DES+BAO+BBN   \cite{Abbott:2017smn}             & $67.2^{+1.2}_{-1.0}$ \\
ACT    \cite{Aiola:2020azj} & $67.9 \pm 1.5$ \\
WMAP9+BAO \cite{Addison:2017fdm}& $68.2 \pm 0.7$ \\
SPT-SZ+BAO  \cite{Addison:2017fdm}  & $68.3 \pm 0.9$ \\
FS+BAO+BBN    \cite{Philcox:2020vvt}& $68.6 \pm 1.1$ \\
\hline 
\cite{Riess:2020fzl}   &  $73.2 \pm 1.3$ \\
\cite{Breuval:2020trd}     &  $73.0 \pm 2.7$ \\
\cite{Riess:2019cxk}     & $74.03 \pm 1.42$\\
\cite{Burns:2018ggj}    &  $73.2 \pm 2.3$\\
\cite{Freedman_2012}    &  $74.3 \pm 2.1$\\
\cite{Soltis:2020gpl}    &  $72.1 \pm 2.0$ \\
\cite{Freedman:2020dne}    &  $69.6 \pm 1.88$ \\
\cite{Yuan:2019npk}    &  $72.4 \pm 2.0$\\
\cite{Jang:2017dxn}    &  $71.17 \pm 2.50$\\
\cite{Reid:2019tiq}   &  $71.1 \pm 1.9$\\
\cite{Huang:2019yhh}    &  $73.3 \pm 4.0$\\
\cite{Pesce:2020xfe}     &  $73.9 \pm 3.0$\\
\cite{Kourkchi:2020iyz}      &  $76.00 \pm 2.55$\\
\cite{Schombert:2020pxm}      &  $75.10 \pm 2.75$ \\
\cite{Blakeslee:2021rqi}     & $73.3 \pm 2.5$\\
\cite{Khetan:2020hmh} & $70.5 \pm 4.1 $ \\
\cite{deJaeger:2020zpb}     &  $75.8^{+5.2}_{-4.9} $\\
\cite{Wong:2019kwg}     &  $73.3 ^{+1.7}_{-1.8}$\\
\cite{Shajib:2019toy}      &  $74.2 ^{+2.7}_{-3.0}$\\
\cite{Birrer:2020tax}     &  $74.5 ^{+5.6}_{-6.1}$ \\
\cite{Birrer:2020tax}     &  $67.4 ^{+4.2}_{-3.2}$\\
\cite{Abbott:2017xzu} & $70_{-8}^{+12}$ \\
\cite{Dominguez:2019jqc} & $67.4_{-6.2}^{+6.0}$\\
\hline 
\end{tabular}
\end{center}
\caption{$H_0$ values at 68\% CL shown in Fig.~\ref{fig-H0}, where we add in quadrature the systematic and statistic errors.}
\label{listH0}
\end{table}

\section{Observational data}
\label{Observations}

To obtain our $H_0$ estimates, that we will use to constrain the cosmological parameters, we combine together most of the recent late time measurements (see Fig.~\ref{fig-H0} and Table~\ref{fig-H0}) with a conservative approach, taking into account that not all measurements are fully independent even between techniques. In particular, we consider the measurements based on:

\begin{enumerate}

\item {\bf Cepheids-SN Ia}: we average the $H_0$ measurements from~\cite{Riess:2020fzl}, $73.2 \pm 1.3$ km/s/Mpc at 68\% CL, from~\cite{Breuval:2020trd} with $H_0 = 73.0 \pm 2.7$ km/s/Mpc at $68\%$ CL, from~\cite{Riess:2019cxk}, $H_0 = 74.03 \pm 1.42$ km/s/Mpc at 68\% CL, from~\cite{Burns:2018ggj}, $H_0 = 73.2 \pm 2.3$ km/s/Mpc at 68\% CL, from~\cite{Freedman_2012}, $H_0 = 74.3 \pm 2.1$ km/s/Mpc at 68\% CL, and we use the smallest error bar of the group.~\footnote{ We compute the arithmetic average of these five measurements, i.e. $\frac{(73.2+73.0+74.03+73.2+74.3)}{5}=73.55$ km/s/Mpc, but instead of quoting the error derived by the standard deviation (i.e. $0.52$ km/s/Mpc in this case) we adopt a more conservative approach, preferring a larger error, i.e. the smallest error bar of the group, equal to $1.3$ km/s/Mpc. We repeat the same thing for the other cases. It should be noted here that we do not adopt a weighted average because of the correlation of the measurements, but in any case this would give a stronger bound, equal to $73.59\pm0.78$ km/s/Mpc.} Therefore, we obtain $H_0 = 73.55 \pm 1.3$ km/s/Mpc at 68\% CL for the measurements based on the Cepheids-SN Ia.

\item {\bf TRGB-SN Ia}: we average the $H_0$ measurements from~\cite{Soltis:2020gpl}, $72.1 \pm 2.0$ km/s/Mpc at 68\% CL, from~\cite{Freedman:2020dne}, $H_0 = 69.6 \pm 0.8 (stat) \pm 1.7 (sys)$ km/s/Mpc at 68\% CL, from~\cite{Yuan:2019npk}, $H_0 = 72.4 \pm 2.0$ km/s/Mpc at 68\% CL, from~\cite{Jang:2017dxn}, $H_0 = 71.17 \pm 1.66 (stat) \pm 1.87 (sys)$ km/s/Mpc at 68\% CL, and~\cite{Reid:2019tiq}, $H_0 = 71.1 \pm 1.9$ km/s/Mpc at 68\% CL, and we use the smallest error bar of the group. Therefore, we obtain $H_0 = 71.27 \pm 1.88$ km/s/Mpc at 68\% CL for the measurements based on the Tip of the Red Giant Branch.

\item {\bf Miras-SN Ia}: we consider~\cite{Huang:2019yhh}, i.e. $H_0 = 73.3 \pm 4.0$ km/s/Mpc at 68\% CL for measurement based on the Miras-SN Ia.

\item {\bf Masers}: we consider~\cite{Pesce:2020xfe}, i.e. $H_0 = 73.9 \pm 3.0$ km/s/Mpc at 68\% CL for the measurement based on the Masers.

\item {\bf Tully Fisher}: we average the $H_0$ measurements from Infrared Tully Fisher,~\cite{Kourkchi:2020iyz}, $H_0 = 76.0 \pm 1.1 (stat) \pm 2.3 (sys)$ km/s/Mpc at 68\% CL, and from Baryonic Tully Fisher,~\cite{Schombert:2020pxm}, $H_0 = 75.1 \pm 2.3 (stat) \pm 1.5 (sys)$ km/s/Mpc at 68\% CL, and we use the smallest error bar of the group. Therefore, we obtain $H_0 = 75.55 \pm 2.55$ km/s/Mpc at 68\% CL for the measurements based on the Tully Fisher.

\item {\bf Surface Brightness Fluctuations}: we average the $H_0$ measurements from~\cite{Blakeslee:2021rqi}, $73.3 \pm 0.7 \pm 2.4$ km/s/Mpc at 68\% CL, and from~\cite{Khetan:2020hmh}, $H_0 = 70.50 \pm 2.37(stat)\pm 3.38(sys) $ km/s/Mpc at $68\%$ CL, that uses SN Ia, calibrated on the SBF distance. Therefore, we obtain $H_0 =  71.9 \pm 2.5$ km/s/Mpc at 68\% CL for the measurement based on the Surface Brightness Fluctuations.

\item {\bf SN II}: we consider~\cite{deJaeger:2020zpb}, i.e. $H_0 = 75.8 ^{+5.2}_{-4.9}$ km/s/Mpc at 68\% CL, for the measurement based on the SN II.

\item {\bf Time-delay Lensing}: we average the $H_0$ measurements from~\cite{Wong:2019kwg} for 6 H0LiCOW lenses, $H_0 = 73.3 ^{+1.7}_{-1.8}$ km/s/Mpc at 68\% CL, from~\cite{Shajib:2019toy} for STRIDES, $H_0 = 74.2 ^{+2.7}_{-3.0}$ km/s/Mpc at 68\% CL, from~\cite{Birrer:2020tax} for TDCOSMO (6 H0LiCOW and 1 STRIDES lenses), $H_0 = 74.5 ^{+5.6}_{-6.1}$ km/s/Mpc at 68\% CL, and from~\cite{Birrer:2020tax} for TDCOSMO+SLACS (adding 33 SLACS lenses without time delays), $H_0 = 67.4 ^{+4.2}_{-3.2}$ km/s/Mpc at 68\% CL, and we use the smallest error bar of the group. Therefore, we obtain $H_0 = 72.35 \pm 1.75$ km/s/Mpc at 68\% CL for the measurements based on the Time-delay Lensing.~\footnote{ The results from~\cite{Wong:2019kwg} for 6 H0LiCOW lenses and from~\cite{Shajib:2019toy} for STRIDES were obtained assuming the deflector mass density profiles by either a power-law or stars (constant mass-to-light ratio) plus standard dark matter halos, while the updated analysis in TDCOSMO IV does not rely on these assumptions on the radial density profiles, encodes the mass-sheet degeneracy in the inference and adds external lenses from the SLACS to provide quantitative constraints on the radial density profiles, providing a solid foundation of agnostic assumptions.}

\item{\bf Gravitational Wave Standard Sirens}:  we consider~\cite{Abbott:2017xzu}, i.e. $H_0=70_{-8}^{+12}$ km/s/Mpc at 68\% CL, for the measurement based on the Standard Sirens.

\item{\bf \boldmath $\gamma$-ray Attenuation}: we consider~\cite{Dominguez:2019jqc}, i.e. $H_0=67.4_{-6.2}^{+6.0}$ km/s/Mpc at 68\% CL, for the measurement based on the $\gamma$-ray Attenuation.

\end{enumerate}

At this point we consider a weighted average of the 10 estimates listed above, obtaining our best $H_0$ estimate, i.e. $H_0 = 72.94 \pm 0.75$ km/s/Mpc at $68\%$ CL, in agreement with~\cite{Verde:2019ivm}.
Computing the average over different measurements, made by different teams with different methods, can in principle ensure a more reliable $H_0$ estimate, perhaps canceling possible biases, so we will call our best estimate {\it optimistic}. We notice here that we can safely add all the datasets together because all of them are in agreement within $2\sigma$.
However, there is some overlap between the data (i),(ii), (iii), i.e. Cepheids, TRGB and Miras respectively, and SBF from~\cite{Khetan:2020hmh} that use the same SN Ia ladders, and a more accurate analysis should account for their covariance.
Therefore, in order to check the robustness and consistency of our optimistic estimate, we do a "jackknife test" of the results, producing 10 averages with each leaving out a different dataset, in Table~\ref{jacknife1}, and 45 averages leaving out every combination of two, in Table~\ref{jacknife2}.
In particular, the exclusion of one measurement changes the $H_0$ estimate from the minimum mean value $H_0=72.63$ km/s/Mpc to the maximum mean value $H_0=73.25$ km/s/Mpc (see Table~\ref{jacknife1}), while the exclusion of two measurements changes the $H_0$ estimate from minimum mean value $H_0=72.19$ km/s/Mpc to the maximum mean value $H_0=73.51$ km/s/Mpc (see Table~\ref{jacknife2}).
The robustness test shows that the exclusion of one or two of the listed measurements doesn't change quantitatively our conclusions on the DE models.

Regarding the correlation between the measurements (i),(ii) and (iii) (Cepheids, TRGB and Miras respectively)~\footnote{The measurement from~\cite{Khetan:2020hmh} based on SBF-SN has instead a relative larger uncertainty and is averaged with~\cite{Blakeslee:2021rqi}, that has the galaxies directly in the Hubble flow.} we can see from Table~\ref{jacknife2} that considering only one of them per time does not change significantly our optimistic estimate. In particular we find that removing (i) and (ii) we have $H_0 = 73.1 \pm 1.1$ km/s/Mpc at $68\%$ CL, removing (i) and (iii) we have $H_0 = 72.59 \pm 0.95$ km/s/Mpc at $68\%$ CL, and removing (ii) and (iii) we have $H_0 = 73.25 \pm 0.84$ km/s/Mpc at $68\%$ CL.
For this reason, we make use in the analysis of the DE models of an additional {\it conservative} estimate of $H_0$ from the Table~\ref{jacknife1} excluding one dataset and taking the result with the largest error bar, i.e. $H_0=72.63\pm 0.92$ km/s/Mpc at 68\% CL without the Cepheids-SN Ia based measurement. And we consider an {\it ultra-conservative} estimate from the Table~\ref{jacknife2} excluding two datasets and taking the result with the largest error bar, i.e. $H_0=72.7\pm 1.1$ km/s/Mpc at 68\% CL, without also the Time-Delay Lensing based measurement.

The cosmological constraints on the parameters of the models we are analysing in this paper are then obtained, making use of the following data:

\begin{itemize}
    \item {\bf Planck}: for this dataset we consider the latest temperature and polarization Cosmic Microwave Background (CMB) power spectra data as measured by the final 2018 Planck legacy release (\cite{Aghanim:2018eyx,Aghanim:2019ame}).
    
    \item {\bf ACT+WMAP}: we include this dataset combination as in Ref.~\cite{Aiola:2020azj}, as a crosscheck of the results, making use of the latest temperature and polarization CMB power spectra data as measured by ACT (\cite{Aiola:2020azj}), and combined with WMAP9 (\cite{Bennett:2012zja}) and a $\tau$ prior.
    
    \item {\boldmath $optH_0$}: we use a Gaussian prior on the Hubble constant as obtained by combining together, in an optimistic way, most of the late time measurement, i.e. $H_0 = 72.94 \pm 0.75$ km/s/Mpc at $68\%$ CL. Moreover, we compare our results with those obtained using the conservative prior $H_0=72.63\pm 0.92$ km/s/Mpc at $68\%$ CL, i.e. {\boldmath $consH_0$}, and the ultra-conservative prior $H_0=72.7\pm1.1$ km/s/Mpc at $68\%$ CL, i.e. {\boldmath $ultraH_0$}.
    
\end{itemize}


\begin{table}
\begin{center}
\begin{tabular}{|c|c|c|}
\hline
$H_0$ mean& $H_0$ error                  & excluded \\
\hline
 72.62699   &   0.9233376      &                    1\\

 73.25471    &  0.8215147    &                   2\\
 
72.92314    &  0.7664772              &           3\\

   72.87175   &   0.7776620        &                 4\\

   72.68698    &  0.7878957 &                        5\\
 
   73.03983  &    0.7894195        &              6\\
 
   72.87141    &  0.7612871  &                      7\\
 
   73.06965   &   0.8338738            &             8              \\
   72.95322  &    0.7549235        &       9  \\
     73.02211  &    0.7585800        &      10  \\
\hline 
\end{tabular}
\end{center}
\caption{Robustness test of the results, excluding the measurement indicated in the last column.}
\label{jacknife1}
\end{table}



\begin{table}
\begin{center}
\begin{tabular}{|c|c|c|c|}
\hline
$H_0$ mean& $H_0$ error                  & excluded & excluded  \\
\hline
   73.05838&       1.059989&               1&           2\\
   72.58911 &     0.9489663 &              1 &          3\\
   72.49379  &    0.9704452  &             1  &         4\\
   72.18593   &   0.9905548   &            1   &        5\\
   72.74182    &  0.9935880    &           1    &       6\\
   72.51725     & 0.9391693     &          1     &      7\\
   72.73386      & 1.086945      &         1      &     8\\
   72.64958&      0.9272990       &        1       &    9\\
   72.74957 &     0.9341007        &       1        &  10\\
   73.25271&      0.8394086&               2&           3\\
   73.20239 &     0.8541644 &              2 &          4\\
   72.98889  &    0.8677809  &             2  &         5\\
   73.41869   &   0.8698181   &            2   &        6\\
   73.18552    &  0.8326054    &           2    &       7\\
   73.51043     & 0.9304035     &          2     &      8\\
   73.27682      &0.8243009      &         2      &     9\\
   73.36285&      0.8290675       &        2       &   10\\
   72.85492&      0.7927890&               3 &          4\\
   72.66224 &     0.8036401 &              3  &         5\\
   73.02929  &    0.8052573  &             3   &        6\\
   72.85530   &   0.7754612   &            3    &       7\\
   73.05918    &  0.8526064    &           3     &      8\\
   72.94041     & 0.7687386     &          3      &     9\\
   73.01174      &0.7726005      &         3       &   10\\
   72.59712&      0.8165602&               4 &          5\\
   72.97585 &     0.8182568 &              4  &         6\\
   72.80062  &    0.7870499  &             4   &        7\\
   73.00013   &  0.7800242    &           4     &      9\\
   72.96215    &  0.7840596    &           4     &     10\\
   72.77377&      0.8302036 &              5   &        6\\
   72.60931 &     0.7976639  &             5    &       7\\
   72.77266  &    0.8823864   &            5     &      8\\
   72.70377   &   0.7903527    &           5      &     9\\
   72.77669    &  0.7945514     &          5       &   10\\
   72.97070  &    0.7992452    &           6   &        7\\
   73.21607   &   0.8845285     &          6    &       8\\
   73.05890    &  0.7918909      &         6     &      9\\
   73.13590     & 0.7961143       &        6      &    10\\
   72.99312  &    0.8454798     &          7     &      8\\
   72.88815   &   0.7635028      &         7      &     9\\
   72.95798    &  0.7672859       &        7       &   10\\
   73.09115  &    0.8367882   &            8      &     9\\
   73.17762   &   0.8417761    &           8       &   10\\
   73.03960  &    0.7607720        &       9   &       10\\

\hline 
\end{tabular}
\end{center}
\caption{Robustness test of the results, excluding the measurements indicated in the last 2 columns.}
\label{jacknife2}
\end{table}


\section{Models}
\label{models}

Even if a $\Lambda$CDM model provides a beautiful description of the available cosmological data, we can not ignore the many $H_0$ measurements at late time providing a larger value for the Hubble constant. Our best optimistic estimate of $H_0=72.94\pm0.75$ km/s/Mpc at 68 \% CL is, in fact, at $5.9\sigma$ of disagreement with the Planck estimate in a $\Lambda$CDM model. Moreover, our $H_0$ estimate increases also the tension with the early time solutions, therefore making the late time solutions more appealing. 

Actually, it is well known that BAO, combined with high-z SNe data, constrain the product $r_{\mathrm{drag}}H_0$, and combining them with either CMB or R19 leaves the other choice in tension, so the early time solutions are preferred because can modify both $r_{\mathrm{drag}}$ and $H_0$ in the right directions. However, we have chosen here to explore the most powerful extensions in solving the Hubble tension, without considering BAO and high-z SNe data. Actually, doing so may offer insight on some of the most famous DE models explored in the literature, focusing on just $H_0$ and neglecting for the moment additional datasets, that can bring further systematic errors in the general picture.

Therefore, in this work we analyze three different models, famous for their ability to solve the Hubble constant tension within $2\sigma$, so in good agreement with R19 and with our best optimistic estimate of $H_0$, as well as our conservative and ultra-conservative $H_0$ values.

First of all, we consider the $w$CDM model, where instead of a cosmological constant $w=-1$ there is a dark energy equation of state of the form $w = P/\rho$, where $P$ and $\rho$ are the dark energy pressure and dark energy density respectively, and w is a free parameter time independent. 
    
As a second scenario, we consider the $w_0w_a$CDM model, where there is a dark energy equation of state time dependent following the parametrization independently proposed by~\cite{Chevallier:2000qy} and~\cite{Linder:2002et}, also known as CPL:
\begin{eqnarray}    
w_x(a) = w_0 + w_a (1-a)\label{cpl}.
\end{eqnarray}
In this parametrization, the dark energy equation of state parameter $w_a$ gives the evolution of $w(a)$ with redshift, while $w_0$ gives the value of the dark energy equation of state today.

Finally, we consider an interacting dark energy scenario IDE, i.e. a class of models where the Dark Matter and Dark Energy continuity equations are described by~\cite{Valiviita:2008iv,Gavela:2009cy,Honorez_2010,delCampo:2008jx,Gavela:2010tm,DiValentino:2019ffd,DiValentino:2019jae}:
\begin{eqnarray}
\dot{\rho}_c+3{\cal H}\rho_c &=& Q\,, \\
\dot{\rho}_x+3{\cal H}(1+w)\rho_x &=&-Q\,.
\end{eqnarray}
In these equations, the dot refer to the derivative with respect to the conformal time $\tau$, $\rho_c$ is the dark matter energy density, $\mathcal{H}$ is the conformal expansion rate of the universe, $\rho_x$ is dark energy density. In our analysis, we assume a constant dark energy equation of state $w=-0.999$, and a coupling function $Q$ given by:
\begin{eqnarray}
Q = \xi{\cal H}\rho_x\,,
\end{eqnarray}
where $\xi$ is a negative dimensionless parameter, describing the coupling between the dark matter and the dark energy fluids.

\begin{table}
\begin{center}
\begin{tabular}{|c|c|c|}
\hline
Parameter                    & prior \\
\hline
$\Omega_{\rm b} h^2$         & $[0.005,0.1]$ \\
$\Omega_{\rm c} h^2$         & $[0.001,0.99]$ \\
$100\theta_{MC}$             & $[0.5,10]$ \\
$\tau$                       & $[0.01,0.8]$ \\
$n_\mathrm{S}$               & $[0.7,1.3]$ \\
$\log[10^{10}A_{s}]$         & $[1.7, 5.0]$ \\
$w_0$                        & $[-3,1]$ \\
$w_a$                        & $[-3,2]$ \\
$\xi$                        & $[-1,0]$ \\
\hline 
\end{tabular}
\end{center}
\caption{Flat priors on the cosmological parameters varied in this work.}
\label{priors}
\end{table}

\section{Methodology}
\label{Method}

We analyse three different baseline models, namely $w$CDM, $w_0w_a$CDM and IDE described in Section~\ref{models}. The common 6 cosmological parameters of the models considered in this work are the baryon energy density $\Omega_{\rm b}h^2$, the cold dark matter energy density $\Omega_{\rm c}h^2$, the ratio between the sound horizon and the 
angular diameter distance at decoupling $\theta_{MC}$, the reionization optical depth 
$\tau$, the amplitude of the scalar primordial power spectrum 
$A_{s}$, the spectral index $n_s$. The specific parameters of the cosmological scenario analysed with the data, are instead the constant Dark Energy equation of state $w$ for the $w$CDM model, the 2 parameters of the redshift dependent Dark Energy equation of state $w_0,w_a$ for the $w_0w_a$CDM model, and the dimensionless coupling $\xi$ for the IDE model. We adopt on the parameters the flat priors listed in Table~\ref{priors}.

For the data analysis, we make use of the publicly available MCMC code \texttt{CosmoMC} (\cite{Lewis:2002ah}) (see \url{http://cosmologist.info/cosmomc/}), also modified to implement the IDE model. This code implements an efficient sampling of the posterior distribution using the fast/slow parameter decorrelations (\cite{Lewis:2013hha}), and makes use of a convergence diagnostic based on the Gelman-Rubin statistics (\cite{Gelman:1992zz}).

\begin{table*}[tb]
\caption{68\% CL constraints for the $w$CDM, $w_0w_a$CDM and IDE scenarios explored in this work, for Planck and Planck+$optH_0$.} 
\label{tab} 
\begin{center}
\resizebox{\textwidth}{!}{  
\begin{tabular}{ c |c c |c c | c c } 
  \hline
 \hline                                          
&  $w$CDM & $w$CDM & $w_0w_a$CDM & $w_0w_a$CDM & IDE & IDE \\ \hline
Parameters & Planck & Planck + $optH_0$ & Planck & Planck + $optH_0$ & Planck & Planck + $optH_0$ \\ \hline

$\Omega_b h^2$ & $    0.02240\pm 0.00015$ & $    0.02238\pm 0.00015$ & $    0.02240 \pm 0.00015$   &    $    0.02240 \pm 0.00015$  & $    0.02239\pm0.000015$ & $    0.02238\pm0.000014$ \\

$\Omega_c h^2$ & $    0.1199\pm0.0014$ & $    0.1201^{+0.0014}_{-0.0015}$  & $    0.1198\pm0.0014$ &   $    0.1198\pm0.0014$  & $    <0.0634$ & $    0.046^{+0.014}_{-0.012}$  \\

$100\theta_{MC}$ & $ 1.04093\pm0.00031$ & $    1.04091\pm0.00031$  & $    1.04094\pm0.00031$  &  $    1.04092\pm0.00031$  & $    1.0458^{+0.0033}_{-0.0021}$ & $    1.0458^{+0.0009}_{-0.0012}$ \\

$\tau$ & $ 0.0540\pm0.0079$ & $    0.0536\pm0.0080$  & $    0.0541^{+0.0073}_{-0.0084}$   & $    0.0544\pm0.0081$   & $    0.0541\pm0.0076$ & $    0.0540\pm0.0076$ \\

$n_s$ & $ 0.9654\pm0.0044$ & $    0.9649^{+0.0052}_{-0.0046}$ & $    0.9657\pm 0.0043$  &   $    0.9655\pm0.0043$  & $    0.9655\pm0.0043$ & $    0.9652\pm0.0042$  \\

${\rm{ln}}(10^{10} A_s)$ & $  3.044\pm0.016$ & $    3.043\pm0.016$  & $    3.044\pm 0.017$   &  $    3.044\pm0.017$   & $    3.044\pm0.0016$ & $    3.044\pm0.0016$ \\

$\xi$ & $    0$ &  $    0$  & $    0$  & $    0$  & $    -0.54^{+0.12}_{-0.28}$ & $    -0.57^{+0.10}_{-0.09}$ \\

$w_0$ & $    -1.58^{+0.16}_{-0.35}$ &  $    -1.187^{+0.038}_{-0.030}$  & $    -1.25^{+0.40}_{-0.56}$  & $    -0.83^{+0.29}_{-0.17}$  & $    -0.999$ & $    -0.999$ \\

$w_a$ & $    0$ &  $    0$  & $    <-0.646$  & $    <-1.05$  & $    0$ & $    0$ \\

\hline

$H_0$ {\rm [km/s/Mpc]} & $    >82.5$ &  $    72.97\pm0.75$  & $    >79.5$  & $    72.96\pm0.74$  & $    72.8^{+3.0}_{-1.5}$ &  $    73.04\pm0.74$\\

$S_8$ & $    0.778 ^{+0.023}_{-0.036}$ &  $    0.817\pm0.015$  & $    0.786^{+0.026}_{-0.044}$  & $    0.819\pm0.015$  & $    1.30^{+0.17}_{-0.44}$ & $    1.24^{+0.09}_{-0.18}$ \\

$r_d$ {\rm [Mpc]}& $    147.08\pm0.30$ &  $    147.05\pm0.31$  & $    147.10\pm0.30$  & $    147.11\pm0.30$  & $    147.08\pm0.30$ &  $    147.06\pm0.29$ \\

\hline \hline 

\end{tabular}
}
\end{center}
\label{table_Planck}
\end{table*}

\begin{table*}[tb]
\caption{68\% CL constraints for the $w$CDM, $w_0w_a$CDM and IDE scenarios explored in this work, for Planck+$consH_0$ and Planck+$ultraH_0$.} 
\label{tab} 
\begin{center}
\resizebox{\textwidth}{!}{  
\begin{tabular}{ c |c c |c c | c c } 
  \hline
 \hline                                          
&  $w$CDM & $w$CDM & $w_0w_a$CDM & $w_0w_a$CDM & IDE & IDE \\ \hline
Parameters & Planck + $consH_0$ & Planck + $ultraH_0$ & Planck + $consH_0$ & Planck + $ultraH_0$ & Planck + $consH_0$ & Planck + $ultraH_0$ \\ \hline

$\xi$ & $    0$ &  $    0$  & $    0$  & $    0$  & $    -0.55\pm0.11$ & $    -0.56\pm0.12$ \\

$w_0$ & $    -1.178^{+0.040}_{-0.033}$ &  $    -1.182^{+0.045}_{-0.038}$  & $    -0.82^{+0.29}_{-0.17}$  & $    -0.82^{+0.29}_{-0.17}$  & $    -0.999$ & $    -0.999$ \\

$w_a$ & $    0$ &  $    0$  & $    <-1.04$  & $    <-1.05$  & $    0$ & $    0$ \\

\hline

$H_0$ {\rm [km/s/Mpc]} & $    72.69\pm 0.91$ &  $    72.8\pm1.1$  & $    72.67\pm 0.92$  & $    72.8\pm1.1$  & $    72.79\pm 0.91$ &  $    72.9\pm1.1$\\

$S_8$ & $    0.818 \pm0.015$ &  $    0.818\pm0.015$  & $    0.820\pm0.015$  & $    0.820\pm0.015$  & $    1.21^{+0.09}_{-0.19}$ & $    1.23^{+0.10}_{-0.22}$ \\

$r_d$ {\rm [Mpc]}& $    147.05\pm0.30$ &  $    147.05\pm0.30$  & $    147.11\pm0.30$  & $    147.11\pm0.31$  & $    147.06\pm0.29$ &  $    147.06\pm0.29$ \\

\hline \hline 

\end{tabular}
}
\end{center}
\label{table_PlanckH0}
\end{table*}

\begin{figure}
\includegraphics[width=0.41\textwidth]{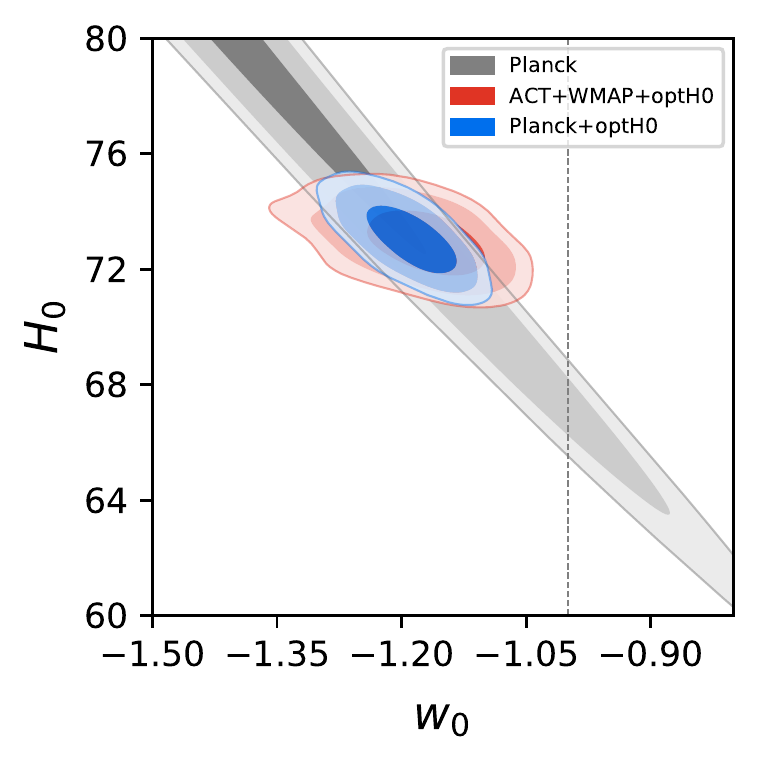}
\caption{68\%, 95\% and 99\% contour plots for the $w$CDM model in the plane ($w$,$H_0$). We can see the evidence for a phantom Dark Energy equation of state at more than $5\sigma$ for Planck+$optH_0$, and at more than $3\sigma$ for ACT+WMAP+$optH_0$.}
\label{fig-w}
\end{figure}

\begin{figure}
\includegraphics[width=0.45\textwidth]{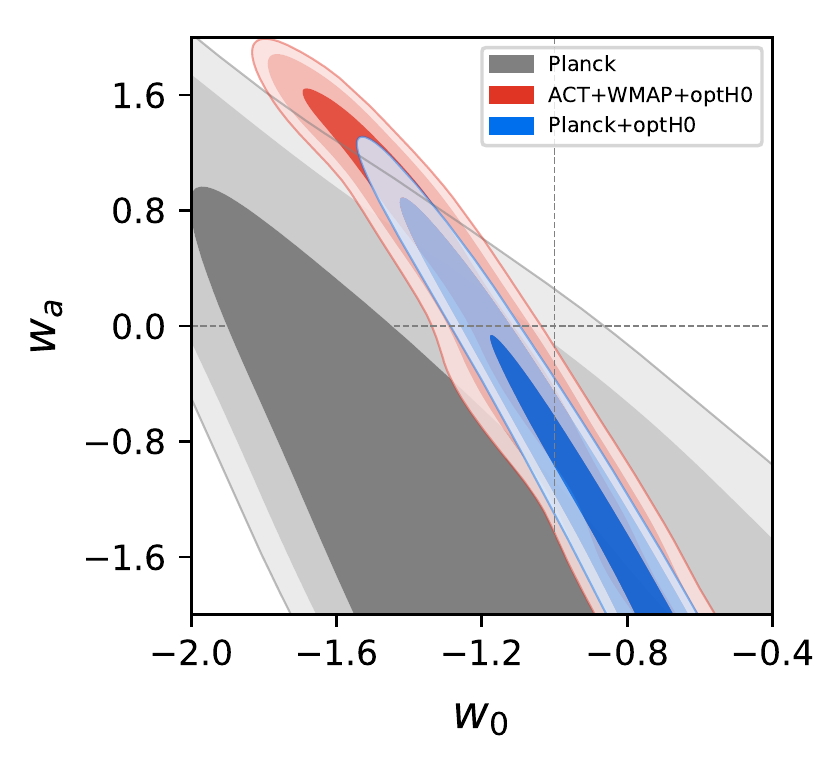}
\caption{68\%, 95\% and 99\% contour plots for the $w_0w_a$CDM model in the plane ($w$,$w_a$). We can see that both the dataset combinations, Planck+$optH_0$ and ACT+WMAP+$optH_0$, are ruling out a cosmological constant, i.e. the point $(w_0=-1,w_a=0)$, at many standard deviations with a large statistical significance.}
\label{fig-w0wa}
\end{figure}

\begin{figure}
\includegraphics[width=0.42\textwidth]{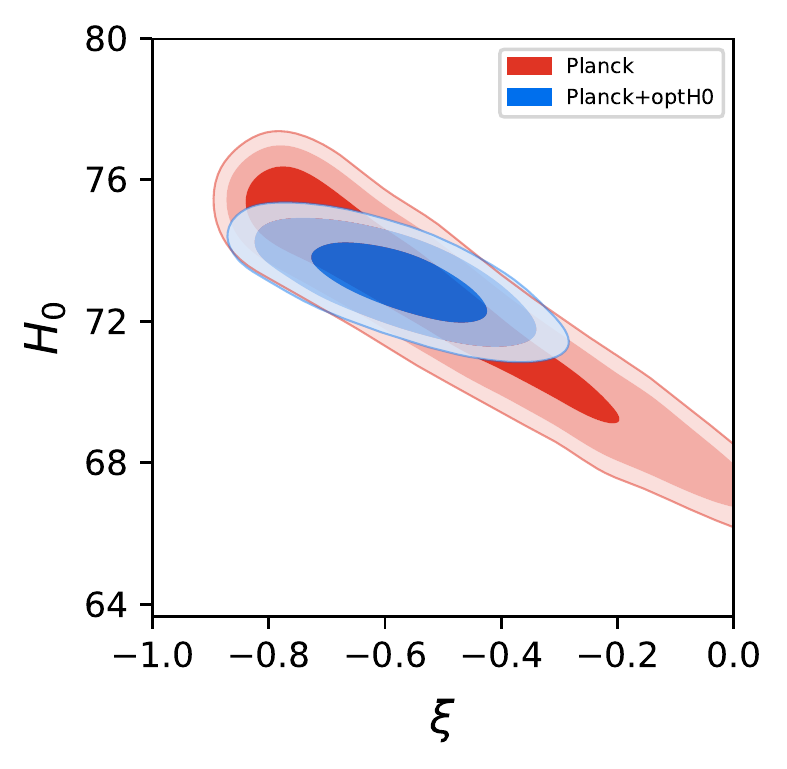}
\caption{68\%, 95\% and 99\% contour plots for the IDE model in the plane ($\xi$,$H_0$). We see for Planck+$optH_0$ a strong evidence for an interaction between the dark energy and the dark matter at more than $6\sigma$.}
\label{fig-xi}
\end{figure}

\begin{table*}[tb]
\caption{68\% CL constraints for the $w$CDM and $w_0w_a$CDM scenarios explored in this work, for ACT+WMAP and ACT+WMAP+$optH_0$.} 
\label{tab} 
\begin{center}
\resizebox{0.82\textwidth}{!}{  
\begin{tabular}{ c |c c |c c } 
  \hline
 \hline                                          
&  $w$CDM & $w$CDM & $w_0w_a$CDM & $w_0w_a$CDM \\ \hline
Parameters & ACT + WMAP & ACT + WMAP + $optH_0$ & ACT + WMAP & ACT + WMAP + $optH_0$ \\ \hline

$\Omega_b h^2$ & $    0.02238\pm 0.00020$ & $    0.02239\pm 0.00020$ & $    0.02238 \pm 0.00021$   &    $    0.02237 \pm 0.00021$   \\

$\Omega_c h^2$ & $    0.1202\pm0.0027$ & $    0.1200\pm0.0025$  & $    0.1203\pm0.0028$ &   $    0.1203^{+0.0026}_{-0.0029}$   \\

$100\theta_{MC}$ & $ 1.04167\pm0.00064$ & $    1.04172\pm0.00066$  & $    1.04168\pm0.00065$  &  $    1.04171^{+0.00070}_{-0.00064}$ \\

$\tau$ & $ 0.062\pm0.013$ & $    0.062\pm0.013$  & $    0.061\pm0.013$   & $    0.061\pm0.013$ \\

$n_s$ & $ 0.9727\pm0.0063$ & $    0.9730\pm 0.0059$ & $    0.9727\pm 0.0064$  &   $    0.9725\pm0.0062$  \\

${\rm{ln}}(10^{10} A_s)$ & $  3.067\pm0.024$ & $    3.066\pm0.024$  & $    3.065\pm 0.024$   &  $    3.064\pm0.024$  \\

$w_0$ & $    -1.12^{+0.56}_{-0.32}$ &  $    -1.172^{+0.052}_{-0.040}$  & $    -0.83^{+0.69}_{-0.83}$ & $    -1.07\pm 0.33$  \\

$w_a$ & $    0$ &  $    0$  & $    <-0.158$  & $    <0.327$  \\

\hline

$H_0$ {\rm [km/s/Mpc]} & $    72^{+9}_{-20}$ &  $    72.87\pm0.73$  & $    70^{+10}_{-20}$  & $    72.90\pm0.75$  \\

$S_8$ & $    0.828 ^{+0.049}_{-0.043}$ &  $    0.827\pm0.025$  & $    0.836\pm0.051$  & $    0.827\pm0.028$   \\

$r_d$ {\rm [Mpc]}& $    147.03\pm0.63$ &  $    147.07\pm0.61$  & $    147.00\pm0.66$  & $    147.02^{+0.68}_{-0.61}$ \\

\hline \hline 

\end{tabular}
}
\end{center}
\label{table_ACT}
\end{table*}

\begin{table*}[tb]
\caption{68\% CL constraints for the $w$CDM and $w_0w_a$CDM scenarios explored in this work, for ACT+WMAP+$consH_0$ and ACT+WMAP+$ultraH_0$.} 
\label{tab} 
\begin{center}
\resizebox{\textwidth}{!}{  
\begin{tabular}{ c |c c |c c } 
  \hline
 \hline                                          
&  $w$CDM & $w$CDM & $w_0w_a$CDM & $w_0w_a$CDM \\ \hline
Parameters & ACT + WMAP + $consH_0$ & ACT + WMAP + $ultraH_0$ & ACT + WMAP + $consH_0$ & ACT + WMAP + $ultraH_0$ \\ \hline

$w_0$ & $    -1.163^{+0.053}_{-0.043}$ &  $    -1.165^{+0.056}_{-0.046}$  & $    -1.06^{+0.38}_{-0.34}$ & $    -1.05^{+0.39}_{-0.33}$  \\

$w_a$ & $    0$ &  $    0$  & $    -0.5^{+1.9}_{-1.3}$  & $    -0.6^{+1.9}_{-1.3}$  \\

\hline

$H_0$ {\rm [km/s/Mpc]} & $    72.58\pm 0.88$ &  $    72.6\pm 1.1$  & $    72.58 \pm 0.92$  & $    72.6\pm1.1$  \\

$S_8$ & $    0.828 \pm0.026$ &  $    0.828\pm0.026$  & $    0.829\pm0.028$  & $    0.828\pm0.028$   \\

$r_d$ {\rm [Mpc]}& $    147.06\pm0.61$ &  $    147.07\pm0.61$  & $    147.01^{+0.68}_{-0.62}$  & $    147.01\pm0.67$ \\

\hline \hline 

\end{tabular}
}
\end{center}
\label{table_ACTH0}
\end{table*}

\section{Results}
\label{results}

We present in Table~\ref{table_Planck} the 68\% CL constraints on the cosmological parameters of the DE models explored in this work ($w$CDM, $w_0w_a$CDM and IDE) obtained combining Planck with our optimistic $H_0$ prior, i.e. $H_0 = 72.94 \pm 0.75$ km/s/Mpc at 68\% CL, presented in Section~\ref{Method}. We show, instead, in Table~\ref{table_ACT} the bounds on the cosmological models $w$CDM and $w_0w_a$CDM we find combining ACT+WMAP and our optimistic $H_0$ prior. We remind here that we can combine these measurements safely because in good agreement with our optimistic $H_0$ prior within these cosmological scenarios.

The $w$CDM cosmological constraints are reported in the first two columns of Table~\ref{table_Planck} and Table~\ref{table_ACT}. We can see a good agreement between Planck and ACT+WMAP, but with Planck preferring a larger Hubble constant and a more phantom dark energy equation of state than ACT+WMAP. In particular, while $H_0>82.5$ km/s/Mpc at 68\% CL for Planck, therefore in agreement with our optimistic $H_0$ estimate at 95\% CL, ACT+WMAP finds $H_0=72^{+9}_{-20}$ km/s/Mpc at 68\% CL, i.e. in agreement within $1\sigma$ with our $optH_0$ prior. Combining these CMB datasets with our optimistic $H_0$, we find instead a perfect agreement between the two dataset combinations, Planck based and ACT based, removing the main differences. Planck+$optH_0$ gives, in fact, an evidence for a phantom Dark Energy equation of state at more than $4.9\sigma$, i.e. $w=-1.187
^{+0.038}_{-0.030}$ at 68\% CL, in agreement with ACT+WMAP+$optH_0$, that finds $w=-1.172^{+0.052}_{-0.040}$ at 68\% CL, ruling out the cosmological constant at more than $3.3\sigma$. The correlation between $w$ and $H_0$ can be seen in Fig.~\ref{fig-w}, as well as the strong evidence for a phantom dark energy coming from both, Planck+$optH_0$ and ACT+WMAP+$optH_0$. To test the robustness of our results, we make use of our conservative $H_0$ estimate $H_0=72.63\pm0.92$ km/s/Mpc at 68\% CL, obtaining $w=-1.178^{+0.040}_{-0.033}$ at 68\% CL for Planck+$consH_0$, corresponding to the first column of Table~\ref{table_PlanckH0}, and $w=-1.163^{+0.053}_{-0.043}$ at 68\% CL for ACT+WMAP+$consH_0$, corresponding to the first column of Table~\ref{table_ACTH0}. Additionally, if we move to our ultra-conservative $H_0$ estimate, $H_0=72.7\pm1.1$ km/s/Mpc at 68\% CL, we obtain $w=-1.182^{+0.045}_{-0.038}$ at 68\% CL for Planck+$ultraH_0$, corresponding to the second column of Table~\ref{table_PlanckH0}, and $w=-1.165^{+0.056}_{-0.046}$ at 68\% CL for ACT+WMAP+$ultraH_0$, corresponding to the second column of Table~\ref{table_ACTH0}. Therefore, we are ruling out the cosmological constant at more than $3\sigma$ in all the cases.

The constraints on the cosmological parameters of the $w_0w_a$CDM scenario are shown in the columns 3 and 4 of Table~\ref{table_Planck} and Table~\ref{table_ACT}. Also in this case, we have a good agreement between Planck and ACT+WMAP, with Planck preferring a larger $H_0$ value than ACT+WMAP. In particular, Planck gives $H_0>79.5$ km/s/Mpc at 68\% CL, in agreement with our $H_0$ estimates at 95\% CL, while ACT+WMAP prefers $H_0=70^{+10}_{-20}$ km/s/Mpc at 68\% CL, in agreement within $1\sigma$ with our $optH_0$ prior. If we now combine these two dataset combinations with our optimistic $H_0$ estimate, Planck+$optH_0$ gives $w_0=-0.83^{+0.29}_{-0.17}$ and $w_a<-1.05$ at 68\% CL, and ACT+WMAP+$optH_0$ finds $w_0=-1.07\pm 0.33$ and $w_a<0.327$ at 68\% CL. In Fig.~\ref{fig-w0wa}, we can see as both, Planck+$optH_0$ and ACT+WMAP+$optH_0$, are ruling out a cosmological constant, i.e. the point $(w_0=-1,w_a=0)$, at more than 3 standard deviations. To check the robustness of our constraints, we make use of our conservative $H_0$ estimate obtaining $w_0=-0.82^{+0.29}_{-0.17}$ and $w_a<-1.04$ at 68\% CL for Planck+$consH_0$, in Table~\ref{table_PlanckH0}, and $w_0=-1.06^{+0.38}_{-0.34}$ and $w_a=-0.5^{+1.9}_{-1.3}$ at 68\% CL for ACT+WMAP+$consH_0$ in Table~\ref{table_ACTH0}. Moreover, for our ultra-conservative $H_0$ estimate, we have $w_0=-0.82^{+0.29}_{-0.17}$ and $w_a<-1.05$ at 68\% CL for Planck+$ultraH_0$ (Table~\ref{table_PlanckH0}) and $w_0=-1.05^{+0.39}_{-0.33}$ and $w_a=-0.6^{+1.9}_{-1.3}$ at 68\% CL for ACT+WMAP+$ultraH_0$ (Table~\ref{table_ACTH0}). In all the cases we are ruling out the cosmological constant in the plane ($w_0,w_a$) at more than $3\sigma$.

The constraints for the IDE scenario are shown in the columns 5 and 6 of Table~\ref{table_Planck}. In this model we have a suspicious evidence for a coupling between the dark matter and the dark energy, possibly due to the correlation of the parameters, as shown in~\cite{DiValentino:2020leo}, for Planck alone. In particular, for this model, Planck gives $H_0=72.8^{+3.0}_{-1.5}$ km/s/Mpc at 68\% CL, in agreement with our optimistic $H_0$ estimate at 68\% CL. If we now combine this dataset with our optimistic $H_0$ prior, we break the degeneracy between the parameters, obtaining for Planck+$optH_0$ a coupling $\xi=-0.57^{+0.10}_{-0.09}$ at 68\% CL, i.e. a strong evidence for an interaction between the dark energy and the dark matter at more than $5.7\sigma$, as we can see also in Fig.~\ref{fig-xi}. It is worthwhile to note that making use of the our conservative $H_0$ prior we have $\xi=-0.55\pm0.11$ at 68\% CL, and of our ultra-conservative $H_0$ prior we have $\xi=-0.56\pm0.12$ at 68\% CL, reducing the evidence for the coupling at $5\sigma$ and $4.7\sigma$ respectively, corresponding to the last two columns of Table~\ref{table_PlanckH0}.

\section{Conclusions}
\label{conclu}

In this paper we study the impact of the Hubble Constant $H_0$ late time measurements on the Dark Energy sector. Firstly, we combine some of the latest $H_0$ measurements, testing the consistency and robustness of the results excluding one, or two, different measurements per time. Then, we define our best optimistic $H_0$ estimate, that is $H_0=72.94\pm0.75$ km/s/Mpc at 68\% CL, obtained averaging over different measurements, made by different teams with different methods, in order to guarantee a more reliable $H_0$ estimate, possibly canceling likely biases.
Finally, we evaluate the impact of this $H_0$ prior on extended Dark Energy cosmologies, in particular $w$CDM, with a constant dark energy equation of state, $w_0w_a$CDM, with a varying with redshift dark energy equation of state, and an IDE scenario, where there is an interaction between dark matter and dark energy. 

We find for $w$CDM that a combination of Planck+$optH_0$ gives an evidence for a phantom Dark Energy equation of state at more than $4.9\sigma$, i.e. $w=-1.187^{+0.038}_{-0.030}$ at 68\% CL, and this result is supported by ACT+WMAP+$optH_0$ that finds $w<-1$ at more than $3\sigma$, i.e. $w=-1.172^{+0.052}_{-0.040}$ at 68\% CL. 

We find for $w_0w_a$CDM that both the dataset combinations, Planck+$optH_0$ and ACT+WMAP+$optH_0$, are ruling out a cosmological constant, i.e. the point $(w_0=-1,w_a=0)$, at more than $3\sigma$.

We see for Planck+$optH_0$ a coupling $\xi=-0.57^{+0.10}_{-0.09}$ at 68\% CL, i.e. a strong evidence for an interaction between the dark energy and the dark matter at more than $5.7\sigma$.

Finally, if we check the robustness of our conclusions making use of a conservative or ultra-conservative $H_0$ priors, we find that these results are confirmed.
We remind here that these DE models are in any case in tension with BAO and high-z SNe data.

\section*{Acknowledgements}
The author thanks Adam Riess for explanations regarding differences in Hubble constant measurements, and Simon Birrer for those based on the Time-delay Lensing. In addition, the author thanks Alessandro Melchiorri and Olga Mena for useful discussions. The author acknowledges the support of the Addison-Wheeler Fellowship awarded by the Institute of Advanced Study at Durham University.

\section*{Data Availability}
We used current publicly available cosmological probes, as listed in the section "Observational Data".

\bibliographystyle{mnras}
\bibliography{H0}

\bsp	
\label{lastpage}
\end{document}